\documentclass[preprint,showpacs,preprintnumbers,amsmath,amssymb]{revtex4}

\usepackage{graphicx}
\usepackage{dcolumn}
\usepackage{bm}
\usepackage{amssymb}

\begin{document}
\newcommand{\vecx}{\mbox{\boldmath $x$}}
\newcommand{\vecp}{\mbox{\boldmath $p$}}

\title{The entropy in finite $N$-unit nonextensive  
systems: The normal average and $q$-average  
}

\author{Hideo Hasegawa
\footnote{hideohasegawa@goo.jp}
}
\affiliation{Department of Physics, Tokyo Gakugei University,  
Koganei, Tokyo 184-8501, Japan}%

\date{\today}

\begin{abstract}
We have discussed the Tsallis entropy in finite $N$-unit nonextensive systems, 
by using the multivariate $q$-Gaussian probability distribution functions (PDFs) 
derived by the maximum entropy methods with the normal average
and the $q$-average ($q$: the entropic index). 
The Tsallis entropy obtained by the $q$-average 
has an exponential $N$ dependence: 
$S_q^{(N)}/N \simeq \:e^{(1-q)N \:S_1^{(1)}} $ 
for large $N$ ($\gg \frac{1}{(1-q)} >0$). 
In contrast, the Tsallis entropy obtained by the normal average 
is given by $S_q^{(N)}/N \simeq [1/(q-1)N]$ for large $N$ ($\gg \frac{1}{(q-1)} > 0)$.
$N$ dependences of the Tsallis entropy
obtained by the $q$- and normal averages are generally quite different, 
although the both results are in fairly good agreement for $\vert q-1 \vert \ll 1.0$.
The validity of the factorization approximation 
to PDFs which has been commonly adopted in the literature, has been examined.
We have calculated correlations defined by 
$C_m= \langle (\delta x_i \:\delta x_j)^m \rangle
-\langle (\delta x_i)^m \rangle\: \langle (\delta x_j)^m \rangle$ 
for $i \neq j$ where $\delta x_i=x_i -\langle  x_i \rangle$, and
the bracket $\langle \cdot \rangle$ stands for the normal and $q$-averages.
The first-order correlation ($m=1$) expresses 
the intrinsic correlation and higher-order correlations 
with $m \geq 2$ include nonextensivity-induced correlation,  
whose physical origin is elucidated in the superstatistics.

\end{abstract}

        

\maketitle
\newpage

\section{Introduction}

In the last decade, much attention has been paid to the nonextensive
statistics since Tsallis proposed the so-called Tsallis entropy
\cite{Tsallis88}-\cite{Tsallis04}.
The Tsallis entropy for $N$-unit nonextensive systems is defined by 
\begin{eqnarray}
S_q^{(N)} &=& \frac{k_B}{1-q}
\left( \int \left[p_q^{(N)}(\vecx)\right]^q \:d \vecx -1 \right),
\label{eq:A1}
\end{eqnarray}
where $q$ is the entropic index, $k_B$ the Boltzmann constant
($k_B=1$ hereafter), $p_q^{(N)}(\vecx)$ denotes the $N$-variate probability
distribution function (PDF), $\vecx = \{ x_i \}$ ($i=1$ to $N$), and 
$d \vecx = \prod_{i=1}^N dx_i $.
The Tsallis entropy is a one-parameter generalization of
the Boltzmann-Gibbs (BG) entropy, to which the Tsallis entropy
reduces in the limit of $q \rightarrow 1.0$,
\begin{eqnarray}
S_{BG}^{(N)} &=& S_1^{(N)}= - \int p_q^{(N)}(\vecx)
\ln p_q^{(N)}(\vecx)\: \:d \vecx.
\end{eqnarray}
The BG entropy is additive because 
for two independent subsystems $A$ and $B$ where the PDF 
is assumed to be given by
\begin{eqnarray}
p_{q}^{(2)}(x_1, x_2) &=& p_q^{(1)}(x_1) \:p_q^{(1)}(x_2)
\hspace{1cm}\mbox{for $x_1 \in A,\; x_2 \in B$},
\label{eq:A11}
\end{eqnarray}
we obtain 
\begin{eqnarray}
S_{1}^{(2)}(A+B) &=& S_1^{(1)}(A)+S_1^{(1)}(B).
\end{eqnarray}
In contrast, the Tsallis entropy is $pseudoadditive$ (non-additive) 
\cite{Tsallis88,Tsallis98,Tsallis01,Tsallis04} because
Eqs. (\ref{eq:A1}) and (\ref{eq:A11}) lead to
\begin{eqnarray}
S_{q,FA}^{(2)}(A+B) &=& S_q^{(1)}(A)+S_q^{(1)}(B) 
+(1-q) S_q^{(1)}(A) \:S_q^{(1)}(B), 
\label{eq:A2}
\end{eqnarray}
where the subscript of FA is attached for a later purpose.
Similarly, when the PDF for $N$-unit independent subsystems is 
assumed to be given in a factorized form,
\begin{eqnarray}
p_{q}^{(N)}(\vecx) &=& \prod_{i=1}^N \: p_q^{(1)}(x_i),
\label{eq:A4}
\end{eqnarray}
we obtain the pseudoadditive Tsallis entropy, $S_{q,FA}^{(N)}$, 
expressed by
\begin{eqnarray}
\ln \left[1+(1-q) S_{q,FA}^{(N)} \right] 
&=& \sum_{i=1}^N \ln \left[1+(1-q) S_{q}^{(1)}(i) \right]. 
\label{eq:D10}
\end{eqnarray}

We should note, however, that Eqs. (\ref{eq:A2}) 
and (\ref{eq:D10}) are not correct in the strict sense 
because the PDF derived by the
maximum-entropy method (MEM) cannot be expressed 
by Eq. (\ref{eq:A11}) or (\ref{eq:A4}), as will be shown shortly 
[Eqs. (\ref{eq:C6})-(\ref{eq:C9})].
Indeed, for identical, independent systems, Eq. (\ref{eq:A1})
with the use of exact multivariate PDFs yields
\begin{eqnarray}
S_{q}^{(2)} &=& 2 S_q^{(1)}+(1-q)\left[S_q^{(1)}\right]^2
+\Delta S_q^{(2)}, 
\label{eq:A12}\\
S_q^{(N)} &=& S_{q,FA}^{(N)} + \Delta S_q^{(N)},
\label{eq:A3}
\end{eqnarray}
where $\Delta S_q^{(N)}$ ($=S_q^{(N)}-S_{q,FA}^{(N)}$)
expresses a correction term. 
Equations (\ref{eq:A12}) and (\ref{eq:A3}) imply that $S_{q}^{(N)}$
does not satisfy the pseudoadditivity because the correction term 
of $\Delta S_q^{(N)}$ is not negligibly small, as will be shown in this study.

It is worthwhile to explain the {\it nonextensivity} which is 
a different concept from the {\it nonadditivity} \cite{Tsallis09}.
An entropy $S^{(N)}$ for $N$-unit system is said to be extensive if 
\begin{eqnarray}
0 < \lim_{N \rightarrow \infty} \;\frac{S^{(N)}}{N} < \infty.
\label{eq:A13}
\end{eqnarray}
For independent (or short-ranged interaction) systems, 
the BG entropy is extensive: $S_{BG}^{(N)}=N S_{BG}^{(1)}$,
while the Tsallis entropy is nonextensive.
In contrast, the BG entropy is nonextensive for systems
with long-ranged interactions. 
The Tsallis entropy may be extensive
for correlated systems such as the probabilistic system \cite{Tsallis05},
some fermionic system \cite{Caruso07},
and Gaussian PDFs with correlation \cite{Wilk07}.

The PDF is commonly evaluated by the MEM 
for the Tsallis entropy with imposing some constraints.
At the moment, there are four possible MEMs: 
(a) original method \cite{Tsallis88},
(b) un-normalized method \cite{Curado91}, 
(c) normalized method \cite{Tsallis98}, and 
(d) the optimal Lagrange multiplier (OLM) method \cite{Martinez00}.
A comparison among the four MEMs is made in Ref. \cite{Tsallis04}.
In relation to the issue on the MEMs, two kinds of definitions
have been considered for an expectation value of physical quantities:
one is the normal average in the MEM (a) and the other is
the $q$-average using the escort probability in the MEMs (c) and (d).
Various arguments have been given that we should employ
the $q$-average \cite{Tsallis04,Abe02,Abe05,Abe06}. 
Recently, however, it has been pointed out that
for a small change of the PDF, thermodynamical averages obtained by
the $q$-average are unstable whereas those obtained by the normal average
are  stable \cite{Abe08,Abe09,Abe09b}. 
In contrast, Ref. \cite{Hanel09} has claimed that for the escort PDF, 
the Tsallis entropy and thermodynamical averages are robust.
This issue on the stability (robustness) of thermodynamical averages 
as well as the Tsallis entropy is currently controversial \cite{Lutsko09}.

In our previous papers \cite{Hasegawa08d,Hasegawa08b}, we discussed
the Tsallis entropy and the generalized Fisher information in nonextensive systems
with and without spatial correlation, calculating the multivariate $q$-Gaussian PDFs. 
The purpose of the present paper is twofold:
(1) to make a comparison between the Tsallis entropies
evaluated by the normal average \cite{Tsallis88}
and the $q$-average \cite{Tsallis98,Martinez00}, 
and (2) to examine the validity of the FA and pseudoadditivity
of the Tsallis entropy.
One of the advantages of a use of the $q$-Gaussian PDF is 
that it is free from an open problem on ambiguity 
in defining the physical temperature
in conformity with the zeroth law of thermodynamics 
\cite{Abe01c}-\cite{Abe06b}.
By using the derived PDFs, we may calculate 
the correlation induced by nonextensivity.
Some related issues of the degree of freedom $N$ on the PDF 
have been discussed in Refs. \cite{Wang02,Wang02b,Jiulin09}.
The nonextensivity-induced correlation has been calculated 
for classical ideal gas \cite{Abe99b,Liyan08,Feng10}
and harmonic oscillator \cite{Liyan08}.
Our calculations will provide some insight to the current controversy 
on the normal versus $q$-averages and show the importance of effects
which are not taken into account in the FA.

The superstatistics is one of alternative approaches to the nonextensive 
statistics besides the MEM \cite{Wilk00,Beck01,Beck05}
(for a recent review, see \cite{Beck07}).
In the superstatistics, it is assumed that {\it locally} the equilibrium state 
of a given system is described by the Boltzmann-Gibbs statistics and
its global properties may be expressed by a superposition over the fluctuating 
intensive parameter ({\it i.e.,} the inverse temperature) 
\cite{Wilk00}-\cite{Beck07}.
The superstatistics has been adopted in many kinds 
of subjects such as hydrodynamic turbulence, 
cosmic ray 
and solar flares \cite{Beck07}. 
The physical origin of the nonextensivity-induced correlation
may be elucidated in the superstatistics.

The paper is organized as follows. In Sec. II, we calculate
the Tsallis entropy, by using  multivariate PDFs for correlated nonextensive systems
derived by the OLM-MEM with the $q$-average 
\cite{Martinez00}\cite{Hasegawa08d,Hasegawa08b}.
In Sec. III, we derive the multivariate PDF by the original MEM \cite{Tsallis88}
with the normal average in order to calculate the Tsallis entropy.
In Sec. IV, we present some numerical calculations of the Tsallis entropy
evaluated by the normal and $q$ averages.
In Sec. V we calculate correlation induced by nonextensivity,
whose physical origin is elucidated in the superstatistics \cite{Wilk00,Beck01}.
Sec. VI is devoted to our conclusion.

\section{OLM-MEM with the $q$-average}
\subsection{$q$-Gaussian PDF}

We consider $N$-unit nonextensive systems whose PDF, $p_q^{(N)}(\vecx)$, 
is derived with the use of the OLM-MEM \cite{Martinez00} for the Tsallis 
entropy given by Eq. (\ref{eq:A1}) \cite{Tsallis88,Tsallis98}.
We impose four constraints given by 
(for details, see Appendix B of Ref. \cite{Hasegawa08b})
\begin{eqnarray}
1 &=& \int p_q^{(N)}(\vecx)\:d \vecx, 
\label{eq:B3}
\\
\mu &=& \frac{1}{N}\sum_{i=1}^N [ x_i ]_q, 
\label{eq:B4}\\
\sigma^2 &=& \frac{1}{N} \sum_{i=1}^N
[ (x_i-\mu)^2 ]_q, \\ 
\label{eq:B5} 
s \:\sigma^2 &=& \frac{1}{N(N-1)} \sum_{i=1}^N \sum_{j=1 (\neq i)}^N
[(x_i-\mu)(x_j-\mu) ]_q.
\label{eq:B6}
\end{eqnarray}
Here $\mu$, $\sigma^2$ and $s$ express the mean, variance, and degree of 
intrinsic correlation, respectively, and the bracket $[ \cdot ]_q$ denotes 
the $q$-average over the escort PDF, $P_q^{(N)}(\vecx)$,
\begin{eqnarray}
[ Q ]_q &=& \int \:P_q^{(N)}(\vecx)\:Q(\vecx) \:d\vecx,
\end{eqnarray}
with
\begin{eqnarray}
P_q^{(N)}(\vecx) &=& \frac{\left(p_q^{(N)}(\vecx) \right)^q} {c_q^{(N)}}, \\
c_q^{(N)} &=& \int \left( p_q^{(N)}(\vecx) \right)^q \:d \vecx,
\label{eq:B7}
\end{eqnarray}
where $Q(\vecx)$ stands for an arbitrary function of $\vecx$.

The OLM-MEM with the constraints given by 
Eqs. (\ref{eq:B3})-(\ref{eq:B6}) leads to the PDF
given by \cite{Hasegawa08b,Note1}
\begin{eqnarray}
p_q^{(N)}(\vecx) &=& \frac{1}{Z_q^{(N)}}
\exp_q\left[- \left( \frac{1}{2 \nu_q^{(N)} \sigma^2 }\right)
\Phi(\vecx)  \right],
\label{eq:C1}
\end{eqnarray}
where
\begin{eqnarray}
\Phi(\vecx) &=& \sum_{i=1}^N \sum_{j=1}^N 
[a_0 \: \delta_{ij}+ a_1 (1-\delta_{ij})]
(x_i-\mu)(x_j-\mu), 
\label{eq:C0} \\
a_0 &=& \frac{[1+(N-2)s]}{(1-s)[1+(N-1)s]},\\
a_1 &=& - \frac{s}{(1-s)[1+(N-1)s]},
\end{eqnarray}
\begin{eqnarray}
Z_q^{(N)} = \left\{ \begin{array}{ll}
r_s^{(N)} \left[\frac{2 \pi \nu_q^{(N)} \sigma^2}{q-1} \right]^{N/2}
\frac{\Gamma\left(\frac{1}{q-1}-\frac{N}{2} \right)}
{\Gamma \left(\frac{1}{q-1} \right)}
\quad & \mbox{for $q > 1 $}, \\ 
r_s^{(N)}(2 \pi \sigma^2)^{N/2}
\quad & \mbox{for $q=1$},  \\
%
r_s^{(N)} \left[\frac{2\pi \nu_q^{(N)} \sigma^2}{1-q}\right]^{N/2}
\frac{\Gamma\left(\frac{1}{1-q}+1 \right)}
{\Gamma \left(\frac{1}{1-q}+\frac{N}{2}+1 \right)}
\quad & \mbox{for $ q <1$},  
\end{array} \right. 
\label{eq:C2}
\end{eqnarray}
\begin{eqnarray}
r_s^{(N)} &=& \{(1-s)^{N-1}[1+(N-1)s]  \}^{1/2}, 
\label{eq:C4}\\
\nu_q^{(N)} &=& \frac{N}{2}(1-q)+1.
\label{eq:C5}
\end{eqnarray}
Here $B(x,y)$ and $\Gamma(z)$ denote the beta and gamma functions, respectively,
and $\exp_q(x)$ expresses the $q$-exponential function defined by
\begin{eqnarray}
\exp_q(x) &=& 
[1+(1-q)x]_{+}^{1/(1-q)},
\label{eq:A8}
\end{eqnarray}
with $[x]_{+} ={\rm max}(x,0)$, which becomes $\exp_q(x) \rightarrow e^x$ 
for $q \rightarrow 1.0$. The entropic index $q$ may take a value,
\begin{eqnarray}
0 < q < 1+\frac{2}{N} \equiv q_U,
\label{eq:C12}
\end{eqnarray}
because $p_q^{(N)}(\vecx)$ given by Eq. (\ref{eq:C1}) has the probability properties
with $\nu_q^{(N)} > 0$ for $q < q_U $ 
and because the Tsallis entropy is stable for $q > 0$ 
\cite{Abe02}.
Evaluations of $Z_q^{(N)}$ and $[ Q ]_q$ with the use 
of the exact approach \cite{Prato95,Rajagopal98} are discussed in Appendix A.1.

In the absence of the intrinsic correlation ($s=0$),
Eq. (\ref{eq:C1}) reduces to \cite{Hasegawa08d}
\begin{eqnarray}
p_q^{(N)}(\vecx) &=& \frac{1}{Z_q^{(N)}}
\exp_q\left[- \left( \frac{1}{2 \nu_q^{(N)} \sigma^2 }\right) 
\sum_{i=1}^N \:(x_i-\mu)^2 \right]. 
\label{eq:C6}
\end{eqnarray}
On the other hand, the PDF in the FA is given by 
%
\begin{eqnarray}
p_{q,FA}^{(N)}(\vecx) &=& \prod_{i=1}^N \:p_q^{(1)}(x_i), \\
&=&  \frac{1}{\left( Z_q^{(1)} \right)^N}  \prod_{i=1}^N 
\exp_q \left[-\left(\frac{1}{2 \nu_q^{(1)} \sigma^2}\right) 
\: (x_i-\mu)^2 \right].
\label{eq:C8}
\end{eqnarray}
In Eqs. (\ref{eq:C6}) and (\ref{eq:C8}), $Z_q^{(N)}$ 
is given by Eq. (\ref{eq:C2}) with $r_s^{(N)}=1.0$.
From a comparison between Eqs. (\ref{eq:C6}) and (\ref{eq:C8}), 
it is evident that
\begin{eqnarray}
p_q^{(N)}(\vecx) & \neq & p_{q,FA}^{(N)}(\vecx),
\label{eq:C9}
\end{eqnarray}
except for $q=1.0$ or $N=1$.

\subsection{Tsallis entropy}

Substituting the PDF given by Eqs. (\ref{eq:C1})-(\ref{eq:C5}) 
to Eq. (\ref{eq:A1}), we obtain the Tsallis entropy 
given by \cite{Hasegawa08d,Hasegawa08b}
\begin{eqnarray} 
S_q^{(N)} &=& \frac{1-c_q^{(N)}}{q-1}, 
\label{eq:D1}
\end{eqnarray}
with
\begin{eqnarray}
c_q^{(N)} &=& \nu_q^{(N)} \left[Z_q^{(N)} \right]^{1-q}.
\label{eq:D2}
\end{eqnarray}
The $s$-dependence of the Tsallis entropy was previously discussed 
(see Fig. 1 of Ref. \cite{Hasegawa08b}).
With increasing $s$, the Tsallis entropy is decreased as given by
\begin{eqnarray}
S_q^{(N)}(s) &=& S_q^{(N)}(0)-\frac{N(N-1)c_q^{(N)}}{4}\:s^2
\hspace{1cm}\mbox{for $\vert s \vert \ll 2/\sqrt{N(N-1)}$ }.
\end{eqnarray}

We pay hereafter our attention to identical, independent systems with $s=0$.
The Tsallis entropy for $q=1.0$ (BG entropy) is given by
\begin{eqnarray}
S_1^{(N)} = N S_1^{(1)},
\end{eqnarray}
with
\begin{eqnarray}
S_1^{(1)}=\left( \frac{1}{2} \right) [\ln(2 \pi \sigma^2)+1]
= 1.418 
\hspace{1cm}\mbox{for $\sigma^2=1.0$}.
\label{eq:D13}
\end{eqnarray}

Employing Eq.(\ref{eq:C8}), we obtain the Tsallis entropy in the FA given by
\begin{eqnarray}
S_{q,FA}^{(N)} &=& \frac{1-(c_q^{(1)} )^N}{q-1},
\label{eq:D5} \\
&=& 
\sum_{k=1}^N \frac{N!}{(N-k)! \:k!}
(1-q)^{k-1} (S_q^{(1)})^k, \\
&=& N S_q^{(1)} + \frac{N(N-1)(1-q)}{2} (S_q^{(1)})^2 + \cdot\cdot.
\label{eq:D11}
\end{eqnarray}
In particular for $N=2$, Eq. (\ref{eq:D11}) becomes
\begin{eqnarray}
S_{q,FA}^{(2)} &=& \frac{1-(c_q^{(1)})^2}{q-1}
=2 S_q^{(1)} +(1-q) (S_q^{(1)})^2. 
\label{eq:D8}
\end{eqnarray}

By using the formula:
\begin{eqnarray}
\ln \left(\frac{\Gamma(z+a)}{\Gamma(z) z^a} \right)
\simeq -\frac{a(1-a)}{2 z}
\hspace{1cm}\mbox{for $\vert z \vert \rightarrow \infty $},
\end{eqnarray}
we obtain the entropy for $\vert q-1 \vert \ll 1.0$ given by
\begin{eqnarray}
S_q^{(N)} &\simeq& S_{q,FA}^{(N)} 
\simeq N S_1^{(1)} -(q-1)\left[\frac{N^2}{2}(S_1^{(1)})^2 - \frac{N}{4} \right]
+\cdot\cdot .
\label{eq:D12}
\end{eqnarray}
The $(q-1)$ term in Eq. (\ref{eq:D12}) includes $O(N^2)$ contributions 
showing the nonadditivity.

For large $N$, we obtain the Tsallis entropy expressed by
\begin{eqnarray}
S_q^{(N)} &\simeq& 
\left(\frac{N}{2}\right) \:e^{(1-q)N \:S_1^{(1)}}, 
\label{eq:P1} \\
S_{q,FA}^{(N)} &\simeq & 
\frac{[\nu_q^{(1)}]^N [Z_q^{(1)}]^{(1-q)N}}{(1-q)}
\hspace{1cm}\mbox{for $N \gg \frac{1}{1-q}> 0$},
\label{eq:P4}
\end{eqnarray}
employing the relation: 
$\ln \Gamma(z) \simeq (z-1/2) \ln z-z +(1/2) \ln 2 \pi +\cdot\cdot$
for $\vert z \vert \rightarrow \infty$.
$S_q^{(N)}$ given by Eq. (\ref{eq:P1}) leads to a very large figure.
In the case of $q=0.5$ and $\sigma^2=1.0$, for example, $S_q^{(N)}/N$ is 
$5.99 \times 10^2$,  $3.09 \times 10^{30}$ and $4.11 \times 10^{307}$ 
for $N=10$, 100 and 1000, respectively.
It yields an astronomical figure for Avogadro's number 
of $N = 6.022 \times 10^{23}$.

\section{Original MEM with the normal average}
\subsection{$q$-Gaussian PDF}
In preceding Sec. II, we have made calculations by using the OLM-MEM
with the $q$-average \cite{Martinez00}. In this section, 
we employ the original MEM
with the normal average \cite{Tsallis88}, imposing the constraints,
\begin{eqnarray}
1 &=& \int \tilde{p}_q^{(N)}(\vecx)\:d \vecx, 
\label{eq:M12} \\
\mu &=& \frac{1}{N}\sum_{i=1}^N \langle x_i \rangle_q, \\
\sigma^2 &=& \frac{1}{N} \sum_{i=1}^N
\langle (x_i-\mu)^2 \rangle_q, \\
s \:\sigma^2 &=& \frac{1}{N(N-1)} \sum_{i=1}^N \sum_{j=1 (\neq i)}^N
\langle (x_i-\mu)(x_j-\mu) \rangle_q,
\label{eq:M13}
\end{eqnarray}
where the bracket $\langle \cdot \rangle_q$ denotes the normal average
(relevant quantities being expressed with tilde hereafter), 
\begin{eqnarray}
\langle Q \rangle_q 
&=& \int \tilde{p}_q^{(N)}(\vecx) \:Q(\vecx) \:d\vecx.
\end{eqnarray}
The original MEM yields \cite{Tsallis88}
\begin{eqnarray}
\tilde{p}_q^{(N)}(\vecx) &=& \frac{1}{\tilde{Z}_q^{(N)}}
{\rm Exp}_{q}\left[- \left( \frac{1}{2 \tilde{\nu}_q^{(N)} \sigma^2}\right) 
\Phi(\vecx) \right],
\label{eq:M1}
\end{eqnarray}
with
\begin{eqnarray}
\tilde{Z}_q^{(N)} = \left\{ \begin{array}{ll}
r_s^{(N)} \left[\frac{2 \pi q \tilde{\nu}_q^{(N)} \sigma^2}{(q-1)} \right]^{N/2}
\frac{\Gamma\left(\frac{1}{q-1}+1 \right)}
{\Gamma \left(\frac{1}{q-1} +\frac{N}{2}+1\right)}
\quad & \mbox{for $q > 1 $}, \\ 
r_s^{(N)}(2 \pi \sigma^2)^{N/2}
\quad & \mbox{for $q=1$},  \\
r_s^{(N)} \left[\frac{2\pi q \tilde{\nu}_q^{(N)} \sigma^2}{(1-q)}\right]^{N/2}
\frac{\Gamma\left(\frac{1}{1-q}-\frac{N}{2} \right)}
{\Gamma \left(\frac{1}{1-q}\right)}
\quad & \mbox{for $ q <1$},  
\end{array} \right. 
\label{eq:M11}
\end{eqnarray}
\begin{eqnarray}
\tilde{\nu}_q^{(N)} &=&  \frac{N}{2}\left(1-\frac{1}{q} \right)+1
= \nu_{1/q}^{(N)}.
\label{eq:M2}
\end{eqnarray}
Here $\Phi(\vecx)$, $r_s^{(N)}$ and $\nu_q^{(N)}$ are given by Eqs. (\ref{eq:C0}),
(\ref{eq:C4}) and (\ref{eq:C5}), respectively, 
and ${\rm Exp}_q(x)$ is the $q$-exponential function defined by \cite{Abe09}
\begin{eqnarray}
{\rm Exp}_q(x) &=& \left[1+\left(1-\frac{1}{q} \right)x \right]_{+}^{1/(q-1)},
\end{eqnarray}
which is different from $\exp_q(x)$ given by Eq. (\ref{eq:A8}).
The two $q$-exponential functions, $\exp_q(x)$ and ${\rm Exp}_q(x)$, 
have the relation \cite{Abe09}:
\begin{eqnarray}
\exp_q(x) &=& {\rm Exp}_{2-q}((2-q)x),\;\;\;\;
{\rm Exp}_q(x) = \exp_{2-q}(x/q).
\end{eqnarray}
The condition of $\tilde{\nu}_q^{(N)} > 0$ implies that
a conceivable $q$ value is 
\begin{eqnarray}
q> 1-\frac{2}{N+2} \equiv q_L=\frac{1}{q_U},
\label{eq:M15} 
\end{eqnarray}
where $q_U$ is given by Eq. (\ref{eq:C12}).
The PDF given by Eqs. (\ref{eq:M1})-(\ref{eq:M2}) is flat-tailed for $q_L < q<1$
and compact support for $q > 1$, which is in contrast with the PDF
given by Eq. (\ref{eq:C6}) obtained by the OLM-MEM.
Evaluations of $\tilde{Z}_q^{(N)}$ and $\langle Q \rangle_q$ with the use 
of the exact approach \cite{Prato95,Rajagopal98} are discussed in Appendix A.2.

\subsection{Tsallis entropy}

By using the PDF given by Eqs. (\ref{eq:M1})-(\ref{eq:M2}), 
we obtain the Tsallis entropy given by
\begin{eqnarray}
\tilde{S}_q^{(N)} &=& \frac{1-\tilde{c}_{q}^{(N)}}{q-1},
\label{eq:M3}
\end{eqnarray}
with
\begin{eqnarray}
\tilde{c}_q^{(N)} &=& \int \left[\tilde{p}_q^{(N)}(\vecx) \right]^q \:d \vecx, 
\label{eq:M14} \\
&=& \frac{\: \left[\tilde{Z}_q^{(N)}\right]^{1-q}}
{\tilde{\nu}_q^{(N)}}.
\label{eq:M4} 
\end{eqnarray}
The Tsallis entropy in the FA is given by
\begin{eqnarray}
\tilde{S}_{q,FA}^{(N)} &=& \frac{1-[\tilde{c}_{q}^{(1)}]^N}{q-1}.
\end{eqnarray}

For $\vert q-1 \vert \ll 1.0$, the entropy is given by
\begin{eqnarray}
\tilde{S}_q^{(N)} &\simeq& \tilde{S}_{q,FA}^{(N)}
\simeq N S_1^{(1)} -(q-1)\left[\frac{N^2}{2}(S_1^{(1)})^2 + \frac{N}{4} \right]
+\cdot\cdot,
\label{eq:M16}
\end{eqnarray}
which is the same as Eq. (\ref{eq:D12}) except for the $O(N)$ term in the bracket.

For large $N$, the Tsallis entropy is expressed by
\begin{eqnarray}
\tilde{S}_q^{(N)} &\simeq& \frac{1}{(q-1)}
\left[1-\frac{2q \:e^{-(q-1)NS_1^{(1)} }}{N(q-1)}   \right],
\label{eq:P2} \\
\tilde{S}_{q,FA}^{(N)} &\simeq& 
\frac{1-[\tilde{\nu}_q^{(1)}]^{-N}[\tilde{Z_q}^{(1)}]^{-(q-1)N}}{q-1}
\hspace{0.5cm}\mbox{for $N \gg \frac{1}{q-1} > 0$},
\label{eq:P3}
\end{eqnarray}
which lead to $\tilde{S}_q^{(N)} \simeq \tilde{S}_{q,FA}^{(N)}$
and $\Delta \tilde{S}_q^{(N)} \simeq 0$ for large $(q-1)N$. 
We note that the $N$- and $q$-dependences of
$\tilde{S}_q^{(N)}$ and $\tilde{S}_{q,FA}^{(N)}$ in the normal average
are quite different from those of $S_q^{(N)}$ [Eq. (\ref{eq:P1})]
and $S_{q,FA}^{(N)}$ [Eq. (\ref{eq:P4})] in the $q$-average.


\section{Numerical calculations}

We show some model calculations in which we hereafter set $\mu=0.0$ and $\sigma^2=1.0$.
Equations (\ref{eq:C12}) and (\ref{eq:M15}) show that the results 
obtained by the $q$- and normal averages
are valid for $0 < q < 1+2/N\equiv q_U$ and $q > 1-2/(N+2)\equiv q_L$, 
respectively (see the inset of Fig. \ref{fig1}).
Although for large $N\vert q-1 \vert$, 
the $N$ dependence of the Tsallis entropy of the $q$-average [Eq. (\ref{eq:P1})]
is quite different from that of $\tilde{S}_q^{(N)}$ of the normal average [Eq. (\ref{eq:P2})],
both entropies are in fairly good agreement for $q_L < q < q_U$.
Figure 1 shows the Tsallis entropy for $N=10$ as a function of $q$
calculated by the $q$-average (solid curve), normal average (chian curve) 
and by the FA in the normal average (dashed curve).
The differences between them are small except for $q \lesssim q_L$ where
$\tilde{S}_q^{(N)}$ divergently increases.

Figures 2(a) and 2(b) show three-dimensional plots of 
$S_q^{(N)}/N$ and $\Delta S_q^{(N)}/S_q^{(N)}$,
respectively, calculated by the $q$-average as functions of $q$ and $N$,
where $\Delta S_q^{(N)}=S_q^{(N)}-S_{q,FA}^{(N)}$.
We note in Fig. 2(a) that $S_q^{(N)}$ is exponentially increased
with increasing $(1-q)$ and/or $N$.
Figure 2(b) shows that $\Delta S_q^{(N)}/S_q^{(N)} \rightarrow 1.0$ 
for large $(1-q) N$, where the Tsallis entropy is given by 
$S_q^{(N)} \cong \Delta S_q^{(N)}$. This implies that
the FA is not a good approximation and
the pseudoadditivity is violated.

Figures 3(a) and 3(b) show three-dimensional plots of $\tilde{S}_q^{(N)}/N$ 
and $\Delta \tilde{S}_q^{(N)}/\tilde{S}_q^{(N)}$, respectively, 
calculated by the normal average as functions of $q$ and $N$, where 
$\Delta \tilde{S}_q^{(N)}=\tilde{S}_q^{(N)}-\tilde{S}_{q,FA}^{(N)}$. 
It is shown that
$\tilde{S}_q^{(N)}$ is decreased with increasing $(q-1)$ and/or $N$, where  
$\Delta \tilde{S}_q^{(N)}/\tilde{S}_q^{(N)}$ is considerably decreased.
The maximum value of $\Delta \tilde{S}_q^{(N)}/\tilde{S}_q^{(N)}$ is about
$0.002$, which means that the pseudoadditivity of $\tilde{S}_q^{(N)}$ 
is approximately satisfied with an accuracy better than 0.998,
and then the FA is a good approximation.

\section{Discussion}

\subsection{Correlations}
\subsubsection{$q$-average}
We will calculate the $m$th-order correlation for $i \neq j$ defined by
\begin{eqnarray}
C_{m} &\equiv & [ (\delta x_i \:\delta x_j)^{m} ]_q
- [ (\delta x_i)^{m} ]_q\: [ (\delta x_i)^{m} ]_q, 
\label{eq:E0}
\end{eqnarray}
which is evaluated by the $q$-average ($\delta x_i=x_i - \mu$). 
With the use of the PDF given 
by Eqs. (\ref{eq:C1})-(\ref{eq:C5}), we obtain the first- and second-order
correlations given by (for details, see Appendix A.1)
\begin{eqnarray}
C_1 &=& [\delta x_i \:\delta x_j]_q
=\sigma^2 s, 
\label{eq:E1}\\
C_2 &=& C_{2s} + C_{2n},
\label{eq:E3}
\end{eqnarray}
with
\begin{eqnarray}
C_{2s} &=& \frac{ 2 [(N+2)-Nq] \:\sigma^4 s^2}
{[(N+4)-(N+2)q ]},
\label{eq:E4}\\
C_{2n} &=& \frac{2(q-1) \:\sigma^4}
{[(N+4)-(N+2)q ]}
\hspace{1cm}\mbox{for $0 < q < 1+ \frac{2}{N+2}$},
\label{eq:E5}
\end{eqnarray}
where we adopt $[\delta x_i]_q=0$.
We note that $C_1$ and $C_{2s}$ arise from intrinsic correlation $s$,
and that $C_{2n}$ expresses correlation induced by nonextensivity 
which vanishes for $q=1.0$ and which approaches
$-2 \sigma^4/N$ as $N \rightarrow \infty$.
In particular for $N=2$, $C_{2s}$ and $C_{2n}$ are given by
\begin{eqnarray}
C_{2s} &=& \frac{2 (2-q) \sigma^4 s^2}{(3-2q) }, \\
C_{2n} &=& \frac{(q-1)\:\sigma^4}{(3-2q) }
\hspace{1cm}\mbox{for $0 < q < \frac{3}{2}$},
\label{eq:E6}
\end{eqnarray}
where  $C_{2s} \geq 0$ for $0 < q < 1.5$ 
while $C_{2n} \leq 0$ for $0 < q \leq 1$
and $C_{2n} > 0$ for $1.0 < q < 1.5$.

On the contrary, the PDF in the FA given by Eq. (\ref{eq:C8}) yields
\begin{eqnarray}
C_1^{FA} &=& C_{2s}^{FA} = C_{2n}^{FA} =0.
\end{eqnarray}

By using Eq. (\ref{eq:C6}), we obtain the $m$th-order correlation
for arbitrary $m$ with $s=0$,
\renewcommand{\arraystretch}{2.0}  
\begin{eqnarray}
C_m 
&=& \left\{ \begin{array}{ll}
A_m^2 \left[\frac{2 \nu_q^{(N)} \sigma^2}{(q-1)} \right]^{m}
\left[\frac{\Gamma\left( \frac{q}{q-1}- \frac{N}{2}-m \right)}
{\Gamma\left( \frac{q}{q-1}- \frac{N}{2} \right)}
-\left(\frac{\Gamma\left( \frac{q}{q-1}- \frac{N}{2} - \frac{m}{2} \right)}
{\Gamma\left( \frac{q}{q-1}- \frac{N}{2} \right)} \right)^2 \right]
\quad & \mbox{for $1.0 < q < q_m$}, \\
A_m^2 \left[\frac{2 \nu_q^{(N)} \sigma^2}{(1-q)} \right]^{m}
\left[\frac{\Gamma\left( \frac{q}{1-q}+ \frac{N}{2}+1 \right)}
{\Gamma\left( \frac{q}{1-q}+ \frac{N}{2} +m+1 \right)} 
-\left(\frac{\Gamma\left( \frac{q}{1-q}+ \frac{N}{2} +1 \right)}
{\Gamma\left( \frac{q}{1-q}+ \frac{N}{2}+ \frac{m}{2} +1 \right)} \right)^2
\right]
\quad & \mbox{for $q <  1.0$}, 
\end{array} \right.  \nonumber \\ 
&& \label{eq:E7}
\end{eqnarray}
where
\renewcommand{\arraystretch}{1.0}
\begin{eqnarray}
A_m &=& \left\{ \begin{array}{ll}
\frac{\Gamma\left(\frac{1}{2}+ \frac{m}{2} \right)}
{\Gamma\left( \frac{1}{2} \right)}
\quad & \mbox{for even $m$}, \\
0
\quad & \mbox{for odd $m$},
\end{array} \right. 
\label{eq:E8} \\
q_m &=& 1 + \frac{2}{N+2(m-1)}.
\end{eqnarray}
The nonextensivity yields higher-order correlation
of $C_m$ with $m \geq 2$ for $q \neq 1.0$ 
in independent nonextensive systems where $s=C_1=0.0$. 

Figure 4 shows $C_2$ as functions of $q$ and $s$ for $N=2$.
At the origin of $(q, s)=(0.0, 0.0)$, we obtain $C_2=-0.33$. 
The parabolic-like line starting from $(q, s)=(1.0, 0.0)$ expresses
the intersection of $C_2$ with the zero-level plane.
With increasing $q$ and/or $s$, $C_2$ is monotonously increased.
As $q$ approaches 1.5, $C_2$ is divergently increased.

Figure 5(a), 5(b) and 5(c) show $C_m$ with $m=2$, 4 and 6, respectively,
as functions of $q$ and $N$.
Magnitudes of $C_m$ are significantly increased with increasing $m$.
$C_6$ shows peculiar $q$ and $N$ dependences.

\subsubsection{Normal average}

We adopt the normal average to evaluate the $m$th-order correlation 
for $j \neq j$ defined by
\begin{eqnarray}
\tilde{C}_{m} &\equiv & \langle (\delta x_i \:\delta x_j)^{m} \rangle_q
-\langle (\delta x_i)^{m} \rangle_q \: \langle (\delta x_i)^{m} \rangle_q.
\end{eqnarray}
With the use of the PDF given by Eqs. (\ref{eq:M1})-(\ref{eq:M2}),
the first- and second-order correlations are given by (for details, see Appendix A.2)
\begin{eqnarray}
\tilde{C}_1 &=& \langle \delta x_i \:\delta x_j \rangle_q
=\sigma^2 s, 
\label{eq:M9}\\
\tilde{C}_2 &=& \tilde{C}_{2s}+\tilde{C}_{2n},
\end{eqnarray}
with
\begin{eqnarray}
\tilde{C}_{2s} &=& \frac{2[(N+2)q-N] \:\sigma^4 \:s^2}{(N+4)q-(N+2)}, 
\label{eq:M8} \\
\tilde{C}_{2n} &=& \frac{2(1-q) \:\sigma^4}{(N+4)q-(N+2)}
\hspace{1cm}\mbox{for $q > 1- \frac{2}{N+4}$}, 
\label{eq:M5}
\end{eqnarray}
where $\langle \delta x_i \rangle_q=0$.
In the case of $N=2$, Eqs. (\ref{eq:M8}) and (\ref{eq:M5}) become 
\begin{eqnarray}
\tilde{C}_{2s} &=& \frac{2(2q-1)\:\sigma^4 \:s^2}{3q-2}, \\
\tilde{C}_{2n} &=& \frac{(1-q) \:\sigma^4}{3q-2}
\hspace{2cm}\mbox{for $q > \frac{2}{3}$}. 
\end{eqnarray}

By using Eq. (\ref{eq:M1}), we obtain correlation
of $\tilde{C}_m$ for arbitrary $m$ with $s=0$,
\renewcommand{\arraystretch}{2.0}  
\begin{eqnarray}
\tilde{C}_m 
&=& \left\{ \begin{array}{ll}
A_m^2 \left[\frac{2 q \tilde{\nu}_q^{(N)} \sigma^2}{(q-1)} \right]^{m}
\left[\frac{\Gamma\left (\frac{1}{q-1}+\frac{N}{2}+1 \right)}
{\Gamma\left(\frac{1}{q-1}+ \frac{N}{2} +m+1 \right)}
-\left(\frac{\Gamma \left( \frac{1}{q-1}+ \frac{N}{2}+1 \right)}
{\Gamma\left(\frac{1}{q-1}+\frac{N}{2} + \frac{m}{2}+1 \right)} \right)^2 \right]
\quad & \mbox{for $q > 1.0$}, \\
A_m^2 \left[\frac{2 q \tilde{\nu}_q^{(N)} \sigma^2}{(1-q)} \right]^{m}
\left[\frac{\Gamma\left( \frac{1}{1-q}- \frac{N}{2} -m \right)}
{\Gamma\left(\frac{1}{1-q}- \frac{N}{2} \right)} 
-\left(\frac{\Gamma\left( \frac{1}{1-q}- \frac{N}{2}-\frac{m}{2} \right)}
{\Gamma\left(\frac{1}{1-q}- \frac{N}{2} \right)} \right)^2
\right]
\quad & \mbox{for $q <  1.0$}, 
\end{array} \right.  \nonumber \\
&& \label{eq:M10}
\end{eqnarray}
where $A_m$ is given by Eq. (\ref{eq:E8}).
When comparing  Eqs. (\ref{eq:M9}), (\ref{eq:M8}), (\ref{eq:M5}) and (\ref{eq:M10}) 
with Eqs. (\ref{eq:E1}), (\ref{eq:E4}), (\ref{eq:E5}) and (\ref{eq:E7}),
respectively, we note that $\tilde{C}_m$ and $C_m$ have the reciprocal symmetry:
$q \leftrightarrow 1/q$.
Then $q$ dependences of $\tilde{C}_m$ are given by those of $C_m$
in Figs. 4 and 5 if we read $q \rightarrow 1/q$.

\subsection{PDF in the superstatistics}
The physical origin of the nonextensivity-induced correlation is easily 
understood in the superstatistics \cite{Wilk00,Beck01,Beck05}.
It is straightforward to apply a concept of the superstatistics to 
composite systems \cite{Wilk07}.
We consider the $N$-unit Langevin model subjected to additive noise 
given by \cite{Hasegawa08b}
\begin{eqnarray}
\frac{dx_i}{dt} &=& -\lambda x_i + \sqrt{2D} \:\xi_i(t)+I
\hspace{1cm}\mbox{for $i=1$ to $N$}, 
\end{eqnarray}
where $ \lambda $ denotes the relaxation rate,
$\xi_i(t)$ the white Gaussian noise with the intensity $D$,
and $I$ an external input.
The PDF of $\pi^{(N)}(\vecx)$ for the system is given by
\begin{eqnarray}
\pi^{(N)}(\vecx) &=& \prod_{i=1}^N \: \pi^{(1)}(x_i),
\label{eq:G0}
\end{eqnarray}
where the univariate PDF of $\pi^{(1)}(x_i)$ obeys the Fokker-Planck equation,
\begin{eqnarray}
\frac{\partial \pi^{(1)}(x_i,t)}{\partial t} 
&=& \frac{\partial }{\partial x_i}[(\lambda x_i -I) \pi^{(1)}(x_i,t)]
+ D \frac{\partial^2 }{\partial x_i^2} \pi^{(1)}(x_i,t).
\end{eqnarray}
The stationary PDF of $\pi^{(1)}(x_i)$ is given by
\begin{eqnarray}
\pi^{(1)}(x_i) &=& \frac{1}{\sqrt{2 \pi \sigma^2}}
\exp\left(-\frac{(x_i- \mu)^2}{2 \sigma^2} \right),
\label{eq:G1}
\end{eqnarray}
with 
\begin{eqnarray}
\mu=I/\lambda, \;\; \sigma^2=D/\lambda.
\end{eqnarray}

After the concept in the superstatistics \cite{Wilk00,Beck01,Beck05,Beck07},
we assume that a model parameter of $\tilde{\beta}$ ($\equiv \lambda/D$)
fluctuates, and that its distribution is expressed by 
the $\chi^2$-distribution with rank $n$ \cite{Wilk00,Beck01}, 
\begin{eqnarray}
f(\tilde{\beta}) &=& 
\frac{1}{\Gamma(n/2)}\left(\frac{n}{2\beta_0} \right)^{n/2}
\tilde{\beta}^{n/2-1} e^{-n \tilde{\beta}/2 \beta_0},
\end{eqnarray}
where $\Gamma(x)$ is the gamma function. Average and variance of $\tilde{\beta}$ 
are given by $ \langle \tilde{\beta} \rangle_{\tilde{\beta}}=\beta_0 $
and $(\langle \tilde{\beta}^2 \rangle_{\tilde{\beta}}-\beta_0^2)/\beta_0^2=2/n$, 
respectively.
Taking the average of $\pi^{(N)}(\vecx)$ over $f(\tilde{\beta})$,
we obtain the stationary PDF given by \cite{Hasegawa08d}
\begin{eqnarray}
p_q^{(N)}(\vecx)&=& \int_0^{\infty} \pi^{(N)}(\vecx) \:f(\tilde{\beta})\:d\tilde{\beta},\\  
& \propto & 
\left[1+\frac{\beta_0}{n}\sum_{i=1}^N (x_i-\mu)^2 \right]^{-(N+n)/2},
\label{eq:G7}
\end{eqnarray}
which is rewritten as
\begin{eqnarray}
p_q^{(N)}(\vecx) & = & \frac{1}{Z_q^{(N)}}
\exp_{q} \left[- \left( \frac{\beta_0}{2 \nu_q^{(N)}} \right) 
\sum_{i=1}^N (x_i-\mu)^2 \right],
\label{eq:G2}
\end{eqnarray}
with
\begin{eqnarray}
Z_q^{(N)} &=& 
\:\left[ \frac{2 \nu_q^{(N)}}{(q-1) \beta_0}\right]^{N/2}
\;\prod_{i=1}^N B\left(\frac{1}{2}, \frac{1}{q-1}-\frac{i}{2} \right), 
\label{eq:G3} 
\\
q &=& 1+ \frac{2}{(N+n)}, 
\label{eq:G5} 
\end{eqnarray}
$\nu_q^{(N)}$ being given by Eq. (\ref{eq:C5}).
In the limit of $n \rightarrow \infty$ ($q \rightarrow 1.0$) where 
$f(\tilde{\beta}) \rightarrow \delta(\tilde{\beta}-\beta_0)$, the PDF reduces to 
the multivariate Gaussian distribution given by
\begin{eqnarray}
p_q^{(N)}(\vecx)  
&= & \frac{1}{(Z_1^{(1)})^N} \prod_{i=1}^N \exp \left[- \frac{\beta_0}{2}\: 
(x_i -\mu)^2 \right],
\label{eq:G6}
\end{eqnarray}
which agrees with Eqs. (\ref{eq:G0}) and (\ref{eq:G1}) 
for $\beta_0=\lambda/D=1/\sigma^2$.

We note that the PDF given by Eq. (\ref{eq:G2}) is equivalent to that
given by Eq. (\ref{eq:C6}) derived by the MEM when we read $\beta_0=1/ \sigma^2$,
although the former is valid for $1 \leq q < 1+2/(N+1)$ 
while the latter for $0 < q < 1+2/N$.
The nonextensivity-induced correlation arises from 
the {\it common} fluctuating field of $\tilde{\beta}$, 
because $p_q^{(N)}(\vecx) \neq \prod_i p_q^{(1)}(x_i)$ for $q \neq 1.0$
despite $\pi^{(N)}(\vecx) = \prod_i \pi^{(1)}(x_i)$. 

Alternatively we may  rewrite the PDF given by Eq. (\ref{eq:G7})
as to conform to the PDF in the normal average for $s=0$
[Eq. (\ref{eq:M1})],
\begin{eqnarray}
p_{\tilde{q}}^{(N)}(\vecx) &\propto& 
{\rm Exp}_{\tilde{q}}\left[- \left(\frac{\beta_0}
{2 \tilde{\nu}_{\tilde{q} }^{(N-2)}} \right)
\sum_{i=1}^N (x_i-\mu)^2 \right],
\label{eq:M6}
\end{eqnarray}
with
\begin{eqnarray}
\tilde{q} &=& 1-\frac{2}{N+n},
\label{eq:M7}
\end{eqnarray}
where $\tilde{\nu}_q^{(N)}$ is given by Eq. (\ref{eq:M2}).
The PDF given by Eqs. (\ref{eq:M6}) and (\ref{eq:M7}) is valid 
for $1-2/(N+n) < \tilde{q} \leq 1$.

\section{Concluding remarks}

\begin{table}
\begin{center}
\caption{A comparison between the Tsallis entropies
obtained by the $q$- and normal averages
[$S_1^{(1)}=(1/2) \{\ln(2 \pi\sigma^2)+1 \}$,
$q_U=1+2/N$ and $q_L = 1-2/(N +2)=1/q_U$].
}
\renewcommand{\arraystretch}{1.5}
\begin{tabular}{|c||c|c|c|} \hline
method
& range &$S_q$ (for $\vert q-1 \vert \ll 1.0$) 
& \;\;$S_q$ (for $N \gg 1$) 
\\ \hline \hline
$q$-average & $0< q < q_U$
& $N S_1^{(1)} - (q-1)\left(\frac{N^2}{2} (S_1^{(1)})^2-\frac{N}{4} \right)$
& $ \frac{N}{2}\:e^{(1-q)NS_1^{(1)}}$  
\\ \hline
normal average & $q > q_L$
& $N S_1^{(1)} - (q-1)\left(\frac{N^2}{2} (S_1^{(1)})^2+\frac{N}{4} \right)$
& $\frac{1}{(q-1)}\left[1-\frac{2q}{N(q-1)} \:e^{-(q-1)NS_1^{(1)}} \right]$ 
\\ \hline
\end{tabular}
\end{center}
\end{table}

Table 1 shows a comparison between the $q$- and normal averages, which
have been obtained with the use of exact $N$-variate $q$-Gaussian PDFs. 
Our calculation has shown the followings:

\noindent
(i) the MEMs with the $q$- and normal averages are valid for
$0 < q < q_U=1+2/N$ and $q > q_L=1- 2/(N+2)$, respectively (see the inset of Fig. 1),

\noindent
(ii) the Tsallis entropy of $S_q^{(N)}$ obtained by the $q$-average \cite{Martinez00}
has an exponential $N$ dependence for large $N$ 
where the FA is not a good approximation and the pseudoadditivity does not hold 
because $\Delta S_q^{(N)}$ $(=S_q^{(N)}-S_{q,FA}^{(N)})$ is considerably large,

\noindent
(iii) the $N$ dependence of $\tilde{S}_q^{(N)}$ for large $N$
obtained by the normal average 
is quite different from that of $S_q^{(N)}$ obtained by the $q$-average,
and the FA in the normal average becomes a good approximation 
because $\Delta \tilde{S}_q^{(N)}$ $(=\tilde{S}_q^{(N)}-\tilde{S}_{q,FA}^{(N)})$ is very small, 

\noindent
(iv) the $q$- and normal averages yield nearly the same results for $q_L < q < q_U$
where the two averages are valid, 

\noindent 
(v) the nonextensivity-induced correlation is realized 
in higher-order correlations ($C_m$ and $\tilde{C}_m$) with $m \geq 2$ 
while the first-order correlation expresses the intrinsic correlation, and

\noindent
(vi) the nonextensivity-induced correlation is elucidated as arising 
from {\it common} fluctuating field introduced in
the superstatistics \cite{Wilk00,Beck01,Beck05}.

\noindent
Items (i)-(iv) clarify the difference and similarity 
between the $q$- and normal averages. Items (ii) and (iii) 
may suggest that the normal average
is more favorable than the $q$-average, which is consistent 
with Refs. \cite{Abe08,Abe09,Abe09b}\cite{Lutsko09}.
We should be careful in employing the FA with the $q$-average
which is shown not to be a good approximating method 
except for $\vert q-1 \vert \ll 1.0$ although it has been widely adopted
in nonextensive classical statistics \cite{Nonext}. 
The situation is the same also in the FA for nonextensive quantum systems 
as recently pointed out in Refs. \cite{Hasegawa09b,Hasegawa09c}.

\begin{acknowledgments}
This work is partly supported by
a Grant-in-Aid for Scientific Research from the Japanese 
Ministry of Education, Culture, Sports, Science and Technology.  
\end{acknowledgments}

\vspace{0.5cm}
\appendix*

\section{A. Evaluations of averages by the exact approach}
\renewcommand{\theequation}{A\arabic{equation}}
\setcounter{equation}{0}

\subsection{The $q$-average}
We first discuss an evaluation of the $q$-average given by
\begin{eqnarray}
Z_q^{(N)}(\alpha) &\equiv &\int \: \left[1-(1-q)  
\alpha \:\Phi(\vecx) \right]^{\frac{1}{1-q}}\:d \vecx, \\
Q_q^{(N)}(\alpha) &=& [ Q(\vecx) ]_q
\equiv \frac{1}{\nu_q^{(N)} Z_q^{(N)}} \int \:Q(\vecx)
\: \left[1-(1-q) \alpha \:\Phi(\vecx) \right]^{\frac{q}{1-q}}\:d \vecx, 
\end{eqnarray}
by using the exact expressions for the gamma function 
\cite{Prato95,Rajagopal98,Hasegawa09b}:
\begin{eqnarray}
y^{-s} &=& \frac{1}{\Gamma(s)} \int_0^{\infty} u^{s-1}e^{-yu}\:du 
\hspace{2cm}\mbox{for $s > 0$}, 
\label{eq:X3}\\
y^s &=&\frac{i}{2 \pi} \Gamma(s+1) \int_C (-t^{-s-1}) e^{-yt}\:dt
\hspace{1cm}\mbox{for $s > 0$}.
\label{eq:X4}
\end{eqnarray}
Here $\alpha=1/(2 \nu_q^{(N)} \sigma^2)$, $\Phi(\vecx)$ is given by
Eq. (\ref{eq:C0}), $Q(\vecx)$ denotes an arbitrary function of $\vecx$,
$C$ the Hankel path in the complex plane, 
and Eq. (\ref{eq:D2}) is employed.
We obtain \cite{Prato95,Rajagopal98,Hasegawa09b}
\renewcommand{\arraystretch}{1.5}
\begin{eqnarray}
Z_q^{(N)}(\alpha)
&=& \left\{ \begin{array}{ll}
\frac{1}{\Gamma\left(\frac{1}{q-1} \right)} 
\int_0^{\infty} u^{\frac{1}{q-1}-1} e^{-u}
Z_1^{(N)}[(q-1) \alpha u] \: du
\quad & \mbox{for $q > 1.0$}, \\
\frac{i}{2 \pi}\Gamma\left(\frac{1}{1-q}+1 \right) 
\int_C (-t)^{-\frac{1}{1-q}-1} e^{-t}
Z_1^{(N)}[-(1-q) \alpha t] \: dt
\quad & \mbox{for $q <  1.0$}, 
\end{array} \right. \nonumber \\
&& \label{eq:X1}
\end{eqnarray}
\begin{eqnarray}
Q_q^{(N)}(\alpha)
&=& \left\{ \begin{array}{ll}
\frac{1}{\nu_q^{(N)} Z_q^{(N)}
\Gamma\left(\frac{q}{q-1} \right)} \int_0^{\infty} u^{\frac{q}{q-1}-1} e^{-u}
Z_1^{(N)} [(q-1) \alpha u] \:Q_1^{(N)}[(q-1) \alpha u] \: du 
\hspace{1cm} & \mbox{for $q > 1.0$}, \\
\frac{i}{2 \pi \nu_q^{(N)} Z_q^{(N)}}
\Gamma\left(\frac{q}{1-q}+1 \right) \int_C (-t)^{- \frac{q}{1-q}-1} e^{-t}
Z_1^{(N)}[-(1-q) \alpha t] \\
\hspace{2cm} \times \:Q_1^{(N)}[-(1-q) \alpha t]\: dt 
\hspace{1cm} & \mbox{for $q <  1.0$}, 
\end{array} \right. \nonumber \\
&& \label{eq:X2}
\end{eqnarray}
where
\begin{eqnarray}
Z_1^{(N)}(\alpha) &=& \int e^{-\alpha \:\Phi(\vecx)}\:d\vecx, 
\label{eq:X5} \\
%
Q_1^{(N)}(\alpha) &=& \frac{1}{Z_1^{(N)}(\alpha)}
\int \:Q(\vecx) \:e^{-\alpha \:\Phi(\vecx)}\:d\vecx.
\label{eq:X6}
\end{eqnarray}
Thus we may evaluate the $q$ average of $Q(\vecx)$ 
from its average over the Gaussian PDF.

For example, simple calculations lead to
\begin{eqnarray}
\langle (x_i-\mu)^2 \rangle_1 &=& \frac{1}{2 \alpha}, 
\label{eq:X10}\\
\langle (x_i-\mu) (x_j-\mu) \rangle_1
&=& \frac{s}{2 \alpha}
\hspace{2cm}\mbox{for $i \neq j$} \\
\langle (x_i-\mu)^2 (x_j-\mu)^2 \rangle_1
&=& \frac{1+ 2 s^2}{4 \alpha^2}
\hspace{1cm}\mbox{for $i \neq j$},\\
\langle (x_i-\mu)^{m} \rangle_1 
&=& \frac{A_m}{\alpha^{m/2}}, \\
\langle (x_i-\mu)^{m} (x_j-\mu)^{m} \rangle_1
&=& \frac{A_m^2}{\alpha^{m}} 
\hspace{1cm}\mbox{for $i \neq j$, $s=0$},
\label{eq:X11}
\end{eqnarray}
with
\renewcommand{\arraystretch}{1.0}
\begin{eqnarray}
A_m &=& \left\{ \begin{array}{ll}
\frac{\Gamma\left( \frac{1}{2}+ \frac{m}{2} \right)}
{\Gamma\left( \frac{1}{2} \right)}
\quad & \mbox{for even $m$}, \\
0
\quad & \mbox{for odd $m$}.
\end{array} \right. 
\end{eqnarray}
Employing Eqs. (\ref{eq:X1}), (\ref{eq:X2}) and (\ref{eq:X10})-(\ref{eq:X11}), 
we obtain 
\begin{eqnarray}
[ (x_i-\mu)^2 ]_q &=& \sigma^2, 
\end{eqnarray}
\begin{eqnarray}
[(x_i-\mu)(x_j-\mu)]_q &=& \sigma^2 s 
\hspace{5cm}\mbox{for $i \neq j$},
\end{eqnarray}
\begin{eqnarray}
[ (x_i-\mu)^2 (x_j-\mu)^2 ]_q
&=& \frac{(N+2-Nq)(1+ 2 s^2) \: \sigma^4}
{(N+4)-(N+2)q}
\hspace{1cm}\mbox{for $i \neq j$}, 
\end{eqnarray}
\renewcommand{\arraystretch}{1.5}
\begin{eqnarray}
[ (x_i-\mu)^{m} ]_q  = \left\{ \begin{array}{ll}
A_m \left[\frac{2 \nu_q^{(N)} \sigma^2}{q-1} \right]^{m/2}
\frac{\Gamma\left(\frac{q}{q-1}-\frac{N}{2}-\frac{m}{2} \right)}
{\Gamma \left(\frac{q}{q-1} -\frac{N}{2}\right)}
\quad & \mbox{for $q > 1 $}, \\ 
A_m \left[\frac{2 \nu_q^{(N)} \sigma^2}{1-q}\right]^{m/2}
\frac{\Gamma\left(\frac{q}{1-q}+\frac{N}{2}+1 \right)}
{\Gamma \left(\frac{q}{1-q}+\frac{N}{2}+\frac{m}{2}+1\right)}
\quad & \mbox{for $ q <1$},  
\end{array} \right. 
\end{eqnarray}
and for $s=0$ and $i \neq j$,
\begin{eqnarray}
[ (x_i-\mu)^{m} (x_j-\mu)^{m} ]_q  = \left\{ \begin{array}{ll}
A_m^2 \left[\frac{2 \nu_q^{(N)} \sigma^2}{q-1} \right]^{m}
\frac{\Gamma\left(\frac{q}{q-1}-\frac{N}{2}-m \right)}
{\Gamma \left(\frac{q}{q-1} -\frac{N}{2}\right)}
\quad & \mbox{for $q > 1 $}, \\ 
A_m^2 \left[\frac{2 \nu_q^{(N)} \sigma^2}{1-q}\right]^{m}
\frac{\Gamma\left(\frac{q}{1-q}+\frac{N}{2}+1 \right)}
{\Gamma \left(\frac{q}{1-q}+\frac{N}{2}+m+1\right)}
\quad & \mbox{for $ q <1$},  
\end{array} \right. 
\end{eqnarray}
which yield Eqs. (\ref{eq:E1}), (\ref{eq:E4}), (\ref{eq:E5}) and (\ref{eq:E7}).

\subsection{The normal average}
Next we discuss an evaluation of the normal average given by
\begin{eqnarray}
\tilde{Z}_q^{(N)}(\alpha) &\equiv&\int \: \left[1-\left(q-1 \right)  
\alpha \:\Phi(\vecx) \right]^{\frac{1}{q-1}}\:d \vecx, \\
\tilde{Q}_q^{(N)}(\alpha) &=& \langle Q(\vecx) \rangle_q
\equiv \frac{1}{\tilde{Z}_q^{(N)}} \int \:Q(\vecx)
\: \left[1-\left(q-1 \right) \alpha \:\Phi(\vecx) \right]^{\frac{1}{q-1}}\:d \vecx,
\end{eqnarray}
where $\alpha=1/(2 q \tilde{\nu}_q^{(N)} \sigma^2)$.
By using Eqs. (\ref{eq:X3}) and (\ref{eq:X4}), we obtain
\begin{eqnarray}
\tilde{Z}_q^{(N)}(\alpha)
&=& \left\{ \begin{array}{ll}
\frac{i}{2 \pi}\Gamma\left(\frac{1}{q-1}+1 \right) 
\int_C (-t)^{-\frac{1}{q-1}-1} e^{-t}
\:Z_1^{(N)}[-(q-1) \alpha t] \: dt
\quad & \mbox{for $q > 1.0$}, \\
\frac{1}{\Gamma\left[1/(1-q) \right]} 
\int_0^{\infty} u^{\frac{1}{1-q}-1} e^{-u}
\:Z_1^{(N)}[(1-q) \alpha u] \: du
\quad & \mbox{for $q <  1.0$}, 
\end{array} \right. \nonumber \label{eq:X7}\\
&& 
\end{eqnarray}
\begin{eqnarray}
&& \tilde{Q}_q^{(N)}(\alpha) \nonumber \\
&=& \left\{ \begin{array}{ll}
\frac{i \:\Gamma[1/(q-1)+1]}{2 \pi \tilde{Z}_q^{(N)}}
\int_C (-t)^{-\frac{1}{q-1}-1} e^{-t}
\:Z_1^{(N)}[-(q-1) \alpha t]\: Q_1[-(q-1) \alpha t]\: dt
\quad & \mbox{for $q > 1.0$}, \\
\frac{1}{\tilde{Z}_q^{(N)} \Gamma\left[1/(1-q) \right]} 
\int_0^{\infty} u^{\frac{1}{1-q}-1} e^{-u}
\:Z_1^{(N)}[(1-q) \alpha u]\: Q_1[(1-q) \alpha t] \: du
\quad & \mbox{for $q <  1.0$}, 
\end{array} \right. \nonumber \label{eq:X8} \\
&& 
\end{eqnarray}
where $Z_1^{(N)}(\alpha)$ and $Q_1^{(N)}(\alpha)$ are given by Eqs. (\ref{eq:X5}) 
and (\ref{eq:X6}), respectively.

Employing Eqs. (\ref{eq:X10})-(\ref{eq:X11}), (\ref{eq:X7}) and (\ref{eq:X8}), 
we obtain
\begin{eqnarray}
\langle (x_i-\mu)^2 \rangle_q &=& \sigma^2, \\
\langle (x_i-\mu) (x_j-\mu) \rangle_q
&=& \sigma^2 s 
\hspace{5cm}\mbox{for $i \neq j$}, \\
\langle (x_i-\mu)^2 (x_j-\mu)^2 \rangle_q
&=& \frac{[(N+2)q-N](1+ 2 s^2) \: \sigma^4}
{(N+4)q-(N+2)}
\hspace{1cm}\mbox{for $i \neq j$},
\end{eqnarray}
\begin{eqnarray}
\langle (x_i-\mu)^{m} \rangle_q  = \left\{ \begin{array}{ll}
A_m \left[\frac{2 q \tilde{\nu}_q^{(N)} \sigma^2}{q-1} \right]^{m/2}
\frac{\Gamma\left(\frac{1}{q-1}+\frac{N}{2}+1 \right)}
{\Gamma \left(\frac{1}{q-1} +\frac{N}{2}+\frac{m}{2}+1\right)}
\quad & \mbox{for $q > 1 $}, \\ 
A_m \left[\frac{2 q \tilde{\nu}_q^{(N)} \sigma^2}{1-q}\right]^{m/2}
\frac{\Gamma\left(\frac{1}{1-q}-\frac{N}{2}-\frac{m}{2} \right)}
{\Gamma \left(\frac{1}{1-q}-\frac{N}{2}\right)}
\quad & \mbox{for $ q <1$},  
\end{array} \right. 
\end{eqnarray}
and for $s=0$ and $i \neq j$,
\begin{eqnarray}
\langle (x_i-\mu)^{m} (x_j-\mu)^{m}\rangle_q  = \left\{ \begin{array}{ll}
A_m^2 \left[\frac{2 q \tilde{\nu}_q^{(N)} \sigma^2}{q-1} \right]^{m}
\frac{\Gamma\left(\frac{1}{q-1}+\frac{N}{2}+1 \right)}
{\Gamma \left(\frac{1}{q-1} +\frac{N}{2}+m+1\right)}
\quad & \mbox{for $q > 1 $}, \\ 
A_m^2 \left[\frac{2 q \tilde{\nu}_q^{(N)} \sigma^2}{1-q}\right]^{m}
\frac{\Gamma\left(\frac{1}{1-q}-\frac{N}{2}-m \right)}
{\Gamma \left(\frac{1}{1-q}-\frac{N}{2}\right)}
\quad & \mbox{for $ q <1$},  
\end{array} \right. 
\end{eqnarray}
which lead to Eqs. (\ref{eq:M9}), (\ref{eq:M8}), (\ref{eq:M5}) and (\ref{eq:M10}).

\newpage

\begin{figure}
\begin{center}
\end{center}
\caption{
(Color online) The $q$-dependence of
$S_q^{(N)}/N$ calculated by the normal average (N-av: solid curve)
and $q$-average (q-av.: chain curve) and in the FA by the
normal average (dashed curve) with $N=10$ for which $q_L=0.833$ and $q_U=1.20$.
The inset shows the validity range of the two averages in the $q$-$N$ space:
the $q$-average and normal average are valid for $0< q < q_U$ and $q_L < q $,
respectively, and for $q_L < q < q_U$ the both averages are valid: 
the dashed line expresses $N=10$.
}
\label{fig1}
\end{figure}

\begin{figure}
\begin{center}
\end{center}
\caption{
(Color online)
(a) $S_q^{(N)}/N$ and (b) $\Delta S_q^{(N)}/S_q^{(N)}$ of the Tsallis entropy
calculated by the $q$ average as functions of $q$ and $N$ ($s=0$): 
note that the ordinate of (a) is in the logarithmic scale.
}
\label{fig2}
\end{figure}

\begin{figure}
\begin{center}
\end{center}
\caption{
(Color online)
(a) $\tilde{S}_q^{(N)}/N$ and (b) $\Delta \tilde{S}_q^{(N)}/\tilde{S}_q^{(N)}$
of the Tsallis entropy calculated by the normal average
as functions of $q$ and $N$ ($s=0$). 
}
\label{fig3}
\end{figure}

\begin{figure}
\begin{center}
\end{center}
\caption{
(Color online)
(a) $C_{2}$ as functions of $q$ and $s$ for $N=2$ obtained by the $q$-average,
the flat plane expressing the zero level.
}
\label{fig4}
\end{figure}

\begin{figure}
\begin{center}
\end{center}
\caption{
(Color online)
(a) $C_{2}$, (b) $C_4$ and (c) $C_6$ as functions of $q$ and $N$ 
obtained by the $q$-average ($s=0.0$).
}
\label{fig5}
\end{figure}

\end{document}